\documentclass[12pt]{emulateapj}
\usepackage{color}

\newcommand\ha{H$\alpha$~}
\newcommand\hb{H$\beta$~}
\newcommand\oii{[O~{\sc ii}]~}
\newcommand\oiii{[O~{\sc iii}]~}
\newcommand\nii{[N~{\sc ii}]~}
\newcommand\pa{P$\alpha$~}

\shorttitle{Nature of \ha selected galaxies . I.}
\shortauthors{Tadaki et al.}

\begin{document}


\title{Nature of \ha selected galaxies at $z>2$. I. Main sequence and dusty star-forming galaxies}


\author{Ken-ichi Tadaki\altaffilmark{1}, Tadayuki Kodama\altaffilmark{1,2}, Ichi Tanaka\altaffilmark{3}, Masao Hayashi\altaffilmark{4}, Yusei Koyama\altaffilmark{1}}
\email{tadaki.ken@nao.ac.jp}

\author{Rhythm Shimakawa\altaffilmark{2}}


\altaffiltext{1}{Optical and Infrared Astronomy Division, National Astronomical Observatory of Japan, Mitaka, Tokyo 181-8588, Japan}
\altaffiltext{2}{Department of Astronomical Science, The Graduate University for Advanced Studies, Mitaka, Tokyo 181-8588, Japan}
\altaffiltext{3}{Subaru Telescope, National Astronomical Observatory of Japan, 650 North A'ohoku Place, Hilo, HI 96720, USA}
\altaffiltext{4}{Institute for Cosmic Ray Research, The University of Tokyo, 5-1-5 Kashiwanoha, Kashiwa, Chiba 277-8582, Japan}


\begin{abstract}
We present the results from our narrow-band imaging surveys of \ha emitters (HAEs) at $z=2.2$ and $z=2.5$
in the Subaru/$XMM$--$Newton$ Deep survey Field with near-infrared camera MOIRCS on the Subaru Telescope.
We have constructed a clean sample of 63 star-forming galaxies at $z=2.2$ and 46 at $z=2.5$.
For 12 (or $\sim$92\%) out of 13 HAEs at $z=2.2$, their \ha emission lines have been
successfully detected by the spectroscopy. 
While about 42\% of the red, massive HAEs with $M_*>10^{10.8} M_\odot$ contain active galactic nuclei (AGNs), most of the blue, less massive ones are likely to be star-forming galaxies. 
This suggests that the AGN may plays an important role in galaxy evolution at the late stage of truncation.
For the HAEs excluding possible AGNs, we estimate the gas-phase metallicities on the basis of [N~{\sc ii}]/H$\alpha$ ratios, and find that the metallicities of the \ha selected
galaxies at $z=2.2$ are lower than those of local star-forming galaxies at fixed stellar mass, as shown by previous studies.
Moreover, we present and discuss the so-called ``main sequence'' of star-forming galaxies at $z>2$ based on our unique sample of HAEs.
By correlating the level of dust extinction with the location on the main sequence,
we find that there are two kinds/modes of dusty star-forming galaxies: star-bursting galaxies and metal-rich normal star-forming galaxies.
\end{abstract}





\section{Introduction}

The star formation rate (SFR) is one of the most fundamental parameters to characterize current activities of galaxies and its time evolution because new stars are continuously formed through cooling and contraction of cold gas in galaxies.
The sketch of the cosmic evolution of star formation activities is known as the ``Madau plot'' \citep{1996MNRAS.283.1388M}.
\cite{2006ApJ...651..142H} have compiled an extensive series of SFR measurements and illustrated the evolution of cosmic SFR density from $z=0$ back to $z\sim6$.
The volume-averaged SFR (i.e.\ SFR density) increases by a factor of 10 from $z = 0$ to $z \sim 1$ and comes to its peak at $z \sim 2$, which indicates that a large fraction of stars in the present-day galaxies were formed beyond $z\sim1$.
The evolution of stellar mass density independently shows that about 50\% of stellar mass in the local galaxies are formed at $z>1$ \citep[e.g][]{2009ApJ...701.1765M,2009ApJ...702.1393K}.
Therefore, star-forming galaxies at the peak epoch are the key population to understanding the formation and early evolution of galaxies.

While the SFR represents a differential rate of on-going star formation,
the stellar mass ($M_*$) indicates the cumulative amount
of past star formation activities.
Therefore, these two are fundamental quantities to characterize star formation
histories of galaxies.
\cite{2007ApJ...670..156D} find a tight correlation between stellar mass and SFR in galaxies at $z\sim2$, named as the ``$main\ sequence$'' of star-forming galaxies.
A similar correlation is also found at $z\sim1$ \citep{2007ApJ...660L..43N} and in the local universe \citep{2007A&A...468...33E}, while the cosmic SFR density declines significantly by more than an order of magnitude from $z=1$ to 0.

In the SFR--$M_*$ diagrams, some outliers have significantly higher specific SFRs (sSFRs) with respect to the main sequence galaxies.
Using the simplified model, \cite{2009MNRAS.398L..58R} has shown that a small difference in sSFR can be dramatically amplified in the subsequent evolution.
While galaxies with high sSFRs undergo a rapid mass accretion, those with low sSFRs evolve with moderate mass increase.
In fact, recent observations of molecular gas in star-forming galaxies at $z=1-2$ have revealed that they are very gas-rich systems, and suggest that two different star formation modes exist; a long-lasting mode for ``normal'' galaxies, and a short-lived intense mode for ``starburst'' galaxies \citep{2010ApJ...714L.118D}.

\begin{figure}
\begin{center}
\includegraphics[scale=1.0]{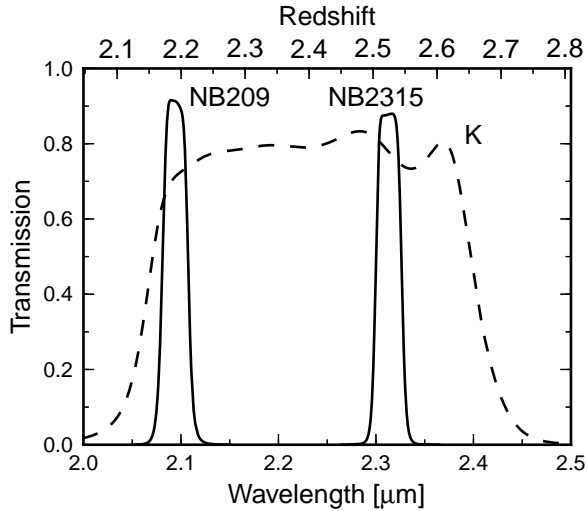}
\caption{The transmission curves of NB209 and NB2315 filters on MOIRCS
(solid line) and the WFCAM $K$-band filter (dashed line). The labels on
the top axis indicate redshifts for HAEs. \label{fig;filter}}
\end{center}
\end{figure}

\cite{2011ApJ...739L..40R} have demonstrated that the main-sequence galaxies represent 98\% of mass-selected star-forming galaxies and account for $\sim$90\% of the cosmic SFR density at $z\sim2$, with the sample detected by PACS 100 $\mu$m or 160 $\mu$m on-board $Herschel$ as well as optical/near-infrared (NIR) color-selected one.
Some recent studies also suggest that the main sequence is independent of the environment in the local universe \citep{2010ApJ...721..193P} and in the distant universe \citep{2013MNRAS.434..423K}, supporting the fact that main sequence galaxies represent a vast majority of star-forming galaxies at each redshift.

Despite its importance, the definition of the main sequence is largely dependent on a sample selection and the SFR indicator used.
In this paper, we present the main sequence of star-forming galaxies at $z>2$, which are selected from a \ha narrow-band (NB) imaging survey.
The advantage of selecting galaxies with NB imaging is that we can construct a nearly SFR-limited, complete sample of star-forming galaxies which spans a broader range in stellar mass from $M_*\sim10^{9}~M_\odot$ to $M_*>10^{11}~M_\odot$.
Also, the \ha line is one of the best SFR indicators
that has many great advantages, such as being less affected by dust extinction, and having been well calibrated in the local universe \citep{2003ApJ...599..971H}.
Recently, we have been conducting the ``$MAHALO-Subaru$'' project (MApping HAlpha and Lines of Oxygen with Subaru; see an overview by \citealt{2013PROCEEDING}), which is a large, systematic survey of \ha emitters (HAEs) utilizing a unique set of NB filters.
We target both biased high-density regions \citep{2011PASJ...63S.415T,2012ApJ...757...15H,2013MNRAS.428.1551K} and non-biased blank fields \citep{2011PASJ...63S.437T}.
In general fields, other large H$\alpha $ surveys have been carried out at $z \sim 2.2$.
\cite{2013MNRAS.428.1128S} have made a large sample of HAEs at $z=2.23$ with a NB filter over $\sim$2 deg$^2$ as a part of the High-$z$ Emission Line Survey (HiZELS) with Wide Field Camera (WFCAM) on UKIRT, and derived the \ha luminosity function and its evolution.
\cite{2010A&A...509L...5H} conducted the deepest survey of HAEs at this high redshift with HAWK-I on VLT over a 56~arcmin$^2$ area, and constrain the faint end slope of the luminosity function. 
Given such large or deep NB surveys providing statistical samples of star-forming galaxies at $z\sim2$, 
our primary motivation is shifted from the stage of identifying high-$z$ star-forming galaxies to the next stage of understanding the physical processes in these galaxies that control their formation and early evolution.

This paper is structured as follows. In Section 2, we describe the existing data and our observations in Subaru/$XMM$--$Newton$ Deep survey Field (SXDF; \citealt{2008ApJS..176....1F}). 
We show the target selections of HAEs in Section 3. 
In Section 4, their global properties such as SFRs, stellar masses, and metallicities are derived, and the mass--metallicity relation is presented.
In Section 5, we discuss the relation between the main-sequence galaxies and dustiness of star formation activities and interpret the mass--metallicity relation observed at $z>2$.
We summarize our study in Section 6.
Throughout this paper, we assume the cosmological parameters of $H_0$=70 km s$^{-1}$ Mpc$^{-1}$, $\Omega_M=0.3$, and $\Omega_\Lambda=0.7$, and Salpeter initial mass function (IMF) is adopted for the estimation of stellar masses and SFRs \citep{1955ApJ...121..161S}.

\section{Data}
\label{sec;survey}

We have conducted the HAE surveys in two redshift slices of $z=2.2$ and $z=2.5$ at a general field SXDF.
Two NB filters, namely NB209 ($\lambda _\mathrm{c}=2.09~\mu$m, FWHM = 0.025~$\mu$m) and NB2315 ($\lambda _\mathrm{c}=2.315~\mu$m, FWHM = 0.027~$\mu$m), are used. 
Figure \ref{fig;filter} shows the filter response functions of the two NB filters together with a broad-band (BB) filter $K$-band which measures the continuum level.
The NB209 and NB2315 filters capture H$\alpha $ emission lines from star-forming galaxies at $z=2.191\pm 0.020$ and $z=2.525\pm 0.021$, respectively. 
Note that $z\sim2.5$ is the highest redshift where we can capture \ha lines with high sensitivity from ground-based telescopes.

\subsection{Broad-band data}
\label{sec;broad-band}

\begin{table}
\begin{center}
\caption{Multi-wavelength data in SXDF.\label{tab;catalog}}
\begin{tabular}{lccc}
\hline
Filter & Instrument & m$_{5\sigma,AB}$ & Reference\\
\hline
$u$ & CFHT/MegaCam & 27.68 & Almaini et al.\\
$B$ & Subaru/Suprime-Cam & 28.38 & \citealt{2008ApJS..176....1F} \\
$V$ & Subaru/Suprime-Cam & 28.01 & \citealt{2008ApJS..176....1F} \\
$R_c$ & Subaru/Suprime-Cam & 27.78 & \citealt{2008ApJS..176....1F} \\
$i'$ & Subaru/Suprime-Cam & 27.69 & \citealt{2008ApJS..176....1F} \\
$z'$ & Subaru/Suprime-Cam & 26.67 & \citealt{2008ApJS..176....1F} \\
$Y$ & VLT/HAWK-I & 26.69 & Fontana et al. \\
$K_s$ & VLT/HAWK-I & 25.92 & Fontana et al.\\
$J$ & UKIRT/WFCAM & 25.63 & \citealt{2007MNRAS.379.1599L} \\
$H$ & UKIRT/WFCAM & 24.76 & \citealt{2007MNRAS.379.1599L} \\
$K$ & UKIRT/WFCAM & 25.39 & \citealt{2007MNRAS.379.1599L} \\
$3.6\mu$m & $Spitzer/$IRAC & 24.72 & \citealt{2013ApJ...769...80A}\\
$4.5\mu$m & $Spitzer/$IRAC & 24.61 & \citealt{2013ApJ...769...80A}\\
$5.8\mu$m & $Spitzer/$IRAC & 22.30 & SpUDS\\
$8.0\mu$m & $Spitzer/$IRAC & 22.26 & SpUDS\\
$24\mu$m & $Spitzer/$MIPS & 30-60 $\mu$Jy & SpUDS\\
\hline
\end{tabular}
\end{center}
\end{table}

In SXDF, there is an extensive data-set covering a wide wavelength from ultraviolet (UV) to mid-infrared.
We use the publicly available multi-wavelength catalog \citep{2013ApJS..206...10G} from the Rainbow Database, which includes the $u$-band data from CFHT/Megacam (O. Almaini et al. in preparation), $B, V , R_c, i'$ and $z'$ band data from Subaru/Suprime-Cam \citep{2008ApJS..176....1F} , $Y$ and $K_s$ band data from VLT/HAWK-I (A. Fontana et al. in preparation), $J, H$ and $K$ bands data from UKIRT/WFCAM (UKIDSS Data Release 8; \citealt{2007MNRAS.379.1599L}), Spitzer/IRAC \citep[SEDS;][]{2013ApJ...769...80A} and MIPS data (SpUDS; PI: J. Dunlop).
Table 1 summarizes the existing BB data that are used in this paper, and their limiting magnitudes.

\subsection{Narrow-band imaging}

\begin{figure}
\begin{center}
\includegraphics[width=1\linewidth]{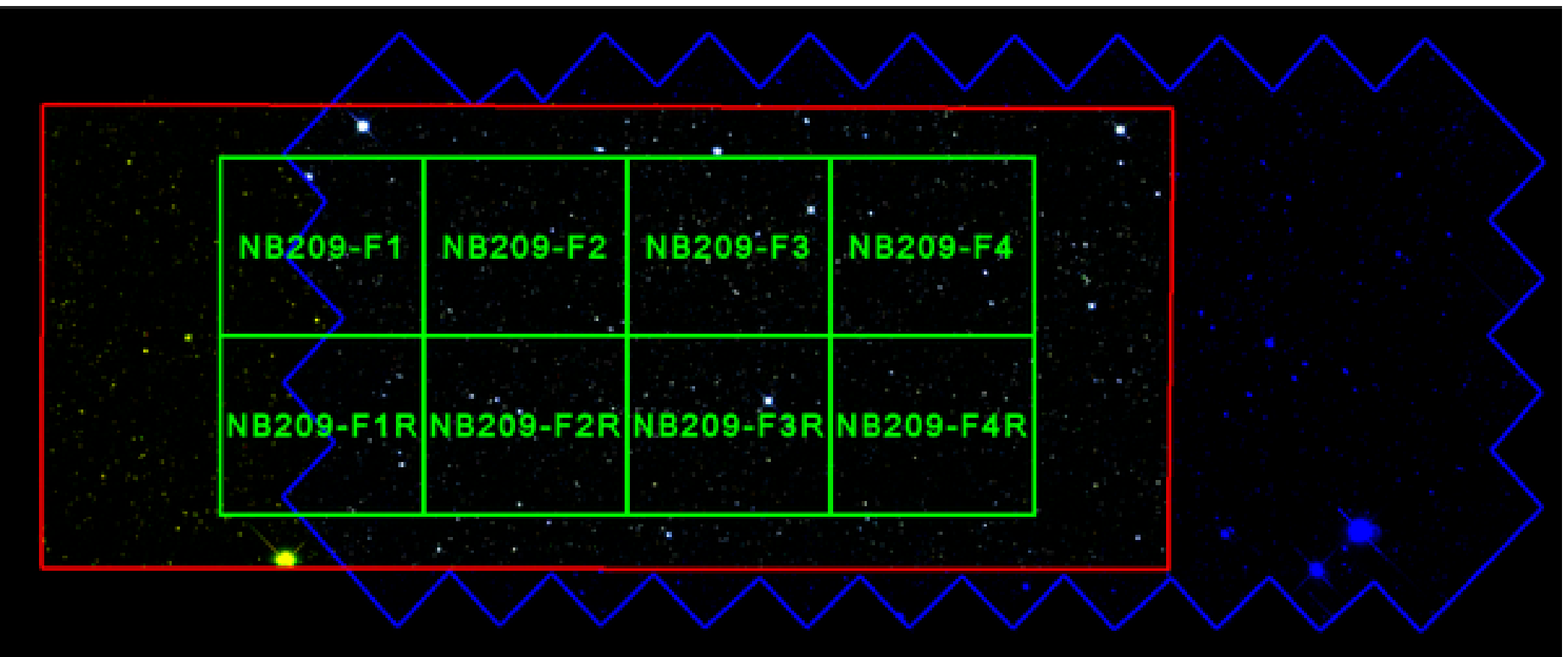}
\includegraphics[width=1\linewidth]{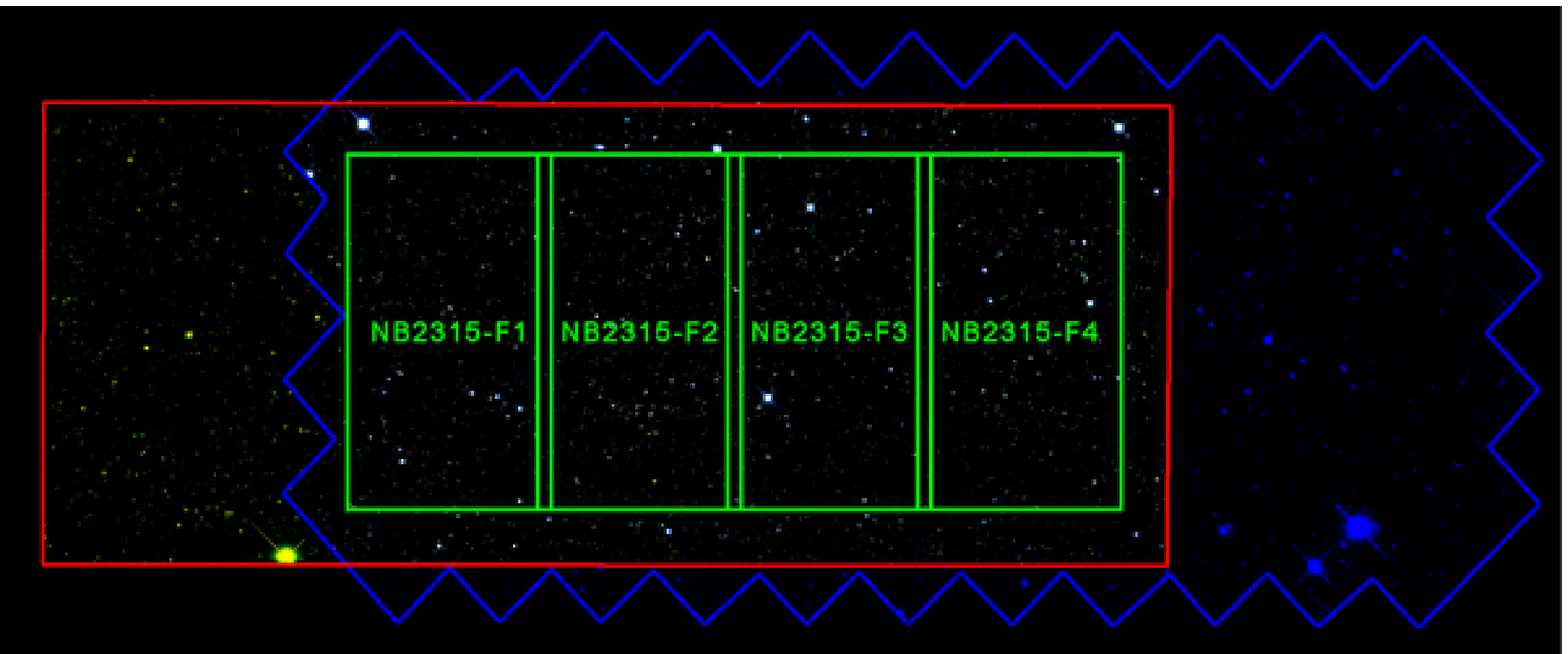} 
\caption{The field coverages of our NB surveys with MOIRCS are shown by green squares. Red squares and blue polygons indicate the areas covered by CANDELS WFC3 and ACS surveys, respectively. \label{fig;survey}}
\end{center}
\end{figure}

The NB imaging observations were carried out with Multi-Object InfraRed Camera and Spectrograph (MOIRCS; \citealt{2008PASJ...60.1347S}) on Subaru.
MOIRCS is equipped with two Hawaii-2 detectors (2048 $\times $2048), and provides a 7\arcmin $\times$ 4\arcmin\ field of view per pointing.
We made observations for seven nights in total, in 2010 October--November and 2011 September.
The weather condition was fine during the observing runs and the seeing sizes were 0.4--0\arcsec.7~in FWHM.
The point spread functions (PSFs) of all the images are eventually smoothed to 0\arcsec.7.
Figure \ref{fig;survey} shows the observed fields, and Table \ref{tab;observation} lists the exposure times, seeing sizes (FWHM), field-of-views (FoVs), and the limiting magnitudes in the NB209 or NB2315 bands. 
The observed field is unique because the high-resolution optical/NIR images by ACS/WFC3 on the $Hubble~Space~Telescope$ are both publicly available \citep{2011ApJS..197...35G}.
The field coverages are slightly different between the NB209 and NB2315 survey.
For NB2315, we spent four MOIRCS pointings which covers a contiguous field of 94 arcmin$^2$.
For NB209, unfortunately, one of the two MOIRCS chips was unavailable at the time of observation due to a mechanical problem.
Therefore, we spent eight MOIRCS pointings to cover a total area of 91 arcmin$^2$.
These correspond to the co-moving volumes of $1.21\times10^4$ Mpc$^3$ and $1.26\times10^4$ Mpc$^3$ at $z=2.2$ (NB209) and $z=2.5$ (NB2315), respectively. 

Data reduction is conducted by using the MOIRCS imaging pipeline software (MCSRED; \citealt{2011PASJ...63S.415T}).
Here, we briefly describe the procedures.
First, a dome flat image is used to correct for any variabilities in sensitivity from pixel to pixel because a sky pattern remains in a self-sky flat image due to interferences of the OH sky emission lines in the case of NB imaging.
The residual pattern, which is estimated by dividing a self-sky flat image by a dome flat image, is subtracted from each flat-fielded image.
Then, a ``median sky'' image, which is made from some adjacent frames before and after the individual image to be processed, is also subtracted from the image. 
After each image is flat-fielded and sky-subtracted, a geometrical distortion is then corrected.
Finally, all the processed images are combined with the inverse square of rms level of each frame as a weight. 
The photometric zero-point of our NB209/NB2315 images are determined using colors of unsaturated stars within the FoVs.

\begin{table}
\begin{center}
\caption{A summary of our NB imaging observations with MOIRCS.\label{tab;observation}}
\begin{tabular}{lcccc}
\hline
Field & Exposure time & Seeing & FoV & $m_{5\sigma}$ \\
&  [min] & [\arcsec] & & [AB,1.6\arcsec] \\
\hline
NB209-F1   & 147 & 0.5 & 4\arcmin $\times$ 3.5\arcmin& 23.5 \\
NB209-F1R &  140 & 0.5 & 4\arcmin $\times$ 3.5\arcmin&23.5 \\
NB209-F2   &  167 & 0.6 & 4\arcmin $\times$ 3.5\arcmin& 23.6 \\
NB209-F2R &  167 & 0.7 & 4\arcmin $\times$ 3.5\arcmin& 23.5 \\
NB209-F3   &  160 & 0.7 & 4\arcmin $\times$ 3.5\arcmin& 23.6 \\
NB209-F3R &  127 & 0.6 & 4\arcmin $\times$ 3.5\arcmin& 23.3 \\
NB209-F4   &  137 & 0.6 & 4\arcmin $\times$ 3.5\arcmin& 23.6 \\
NB209-F4R &  177 & 0.6 & 4\arcmin $\times$ 3.5\arcmin& 23.6 \\
\hline
NB2315-F1 & 180 & 0.6 & 4\arcmin $\times$ 7\arcmin& 22.9\\
NB2315-F2 & 180 & 0.6 & 4\arcmin $\times$ 7\arcmin& 22.9\\
NB2315-F3 & 186 & 0.6 & 4\arcmin $\times$ 7\arcmin& 22.8\\
NB2315-F4 & 186 & 0.6 & 4\arcmin $\times$ 7\arcmin& 22.9\\
\hline
\end{tabular}
\end{center}
\end{table}

\begin{figure*}
\begin{center}
\includegraphics[scale=0.9]{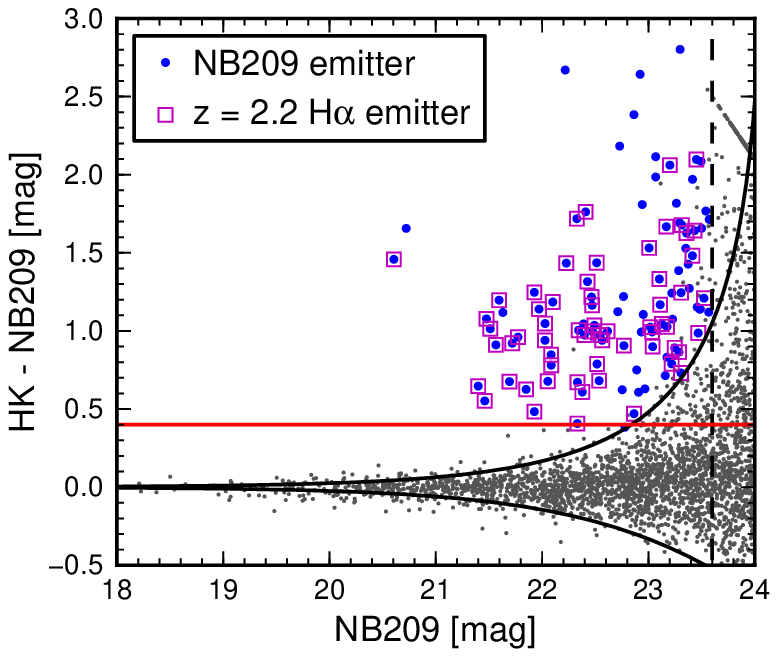}
\includegraphics[scale=0.9]{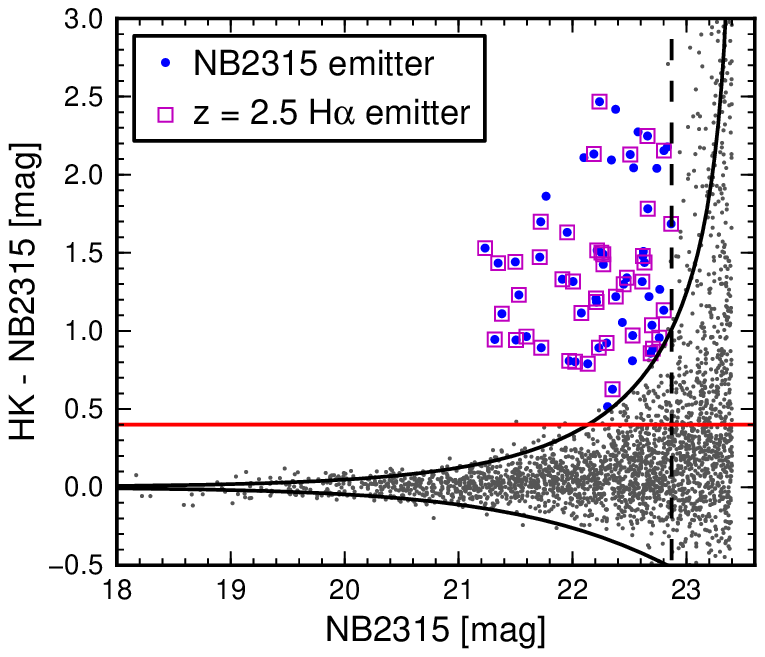}
\caption{$Left$: $HK$ - NB209 color-magnitude diagram to select emission-line galaxies. Grey dots show all the objects detected in the NB209 image and blue circles show the NB emitters which satisfy our selection criteria. The solid curve corresponds to the photometric error of 3$\sigma $, and the red line corresponds to the NB excess of 0.4 magnitude. The vertical dashed line denotes the $5\sigma$ limiting magnitude in the NB209 image.
The HAEs, selected from the NB emitters by the color criteria, are shown by magenta squares (see text for details).  $Right$: The same as the left panel, but for $HK$ - NB2315 color-magnitude diagram. \label{fig;emitter_selection}}
\end{center}
\end{figure*}

\begin{figure*}
\begin{center}
\includegraphics[scale=0.9]{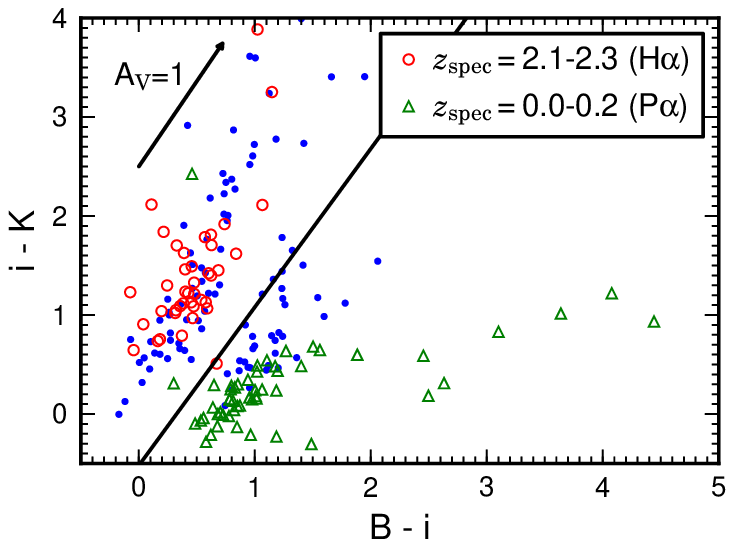}
\includegraphics[scale=0.9]{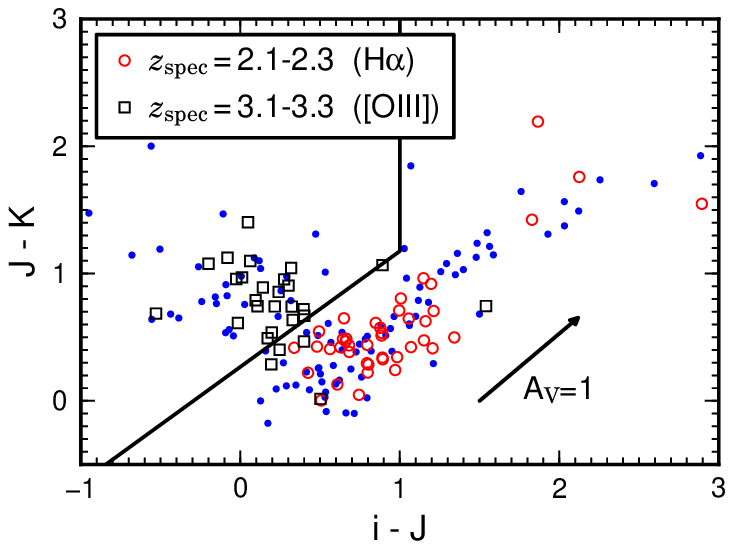}\\
\includegraphics[scale=0.9]{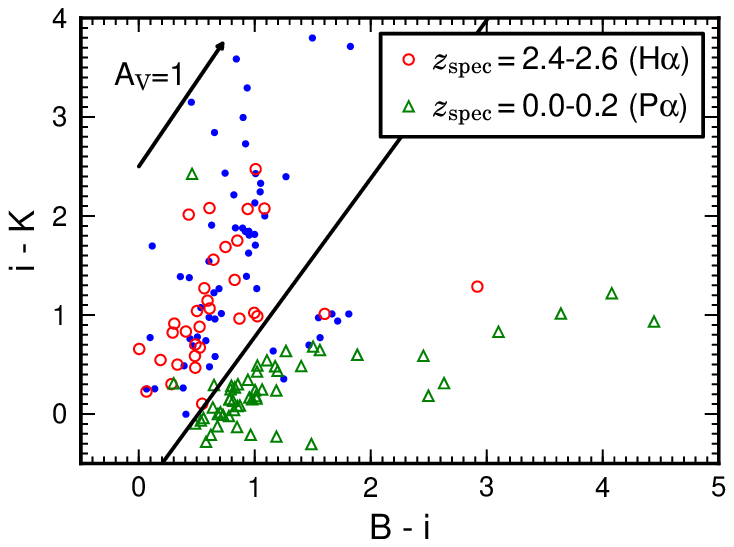}
\includegraphics[scale=0.9]{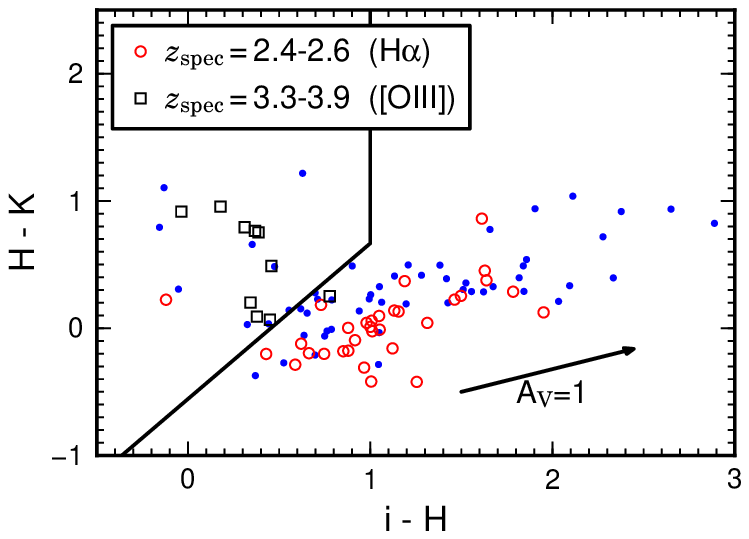}
\caption{$Top\ left$: The $B-i$ versus $i-K$ color--color diagram for the NB209 emitters. $Top\ right$: The $i-J$ versus $J-K$ color--color diagram. Open symbols show the spectroscopic sample of galaxies at $z=2.1$--2.3 (red circles), $z=0.0$--0.2 (green triangles) and $z=3.1$--3.3 (black squares) taken from the MODS catalog \citep{2011PASJ...63S.379K}. Blue dots denote our NB209 emitters. Solid lines indicate the criteria of effectively separating out the HAEs at $z=2.2$. $Bottom\ left$ and $Bottom\ right$: The same as the upper two panels but for the NB2315 emitters. \label{fig;color_selection}}
\end{center}
\end{figure*}

\section{Target selection}
\label{sec;target_selection}

\subsection{Narrow-band emitters}

We identify emission line galaxies on the basis of flux excesses in a NB as compared to a BB that covers the NB wavelength.
First, the NB detected catalogs are made from the images ($H,K$ from WFCAM and NB from MOIRCS) using SExtractor \citep{1996A&AS..117..393B}.
The pixel scales and PSF sizes of the BB images are matched to those of the NB images. The offset between these images is smaller than 0.05 arcsec. 
Source detections and photometries are carried out using the double image mode of the SExtractor. The extraction criterion is having at least 9 pixels with fluxes above 1.5$\sigma$ level in the NB images. The sources fainter than 5$\sigma$ limiting magnitudes in the detection frame (NB209 or NB2315) are rejected.
An aperture magnitude within a diameter of 1\arcsec.6 is used to derive a color index.
Note that these catalogs are used for the selection of NB emitters and the measurements of NB fluxes.
For the following color selection (Figure \ref{fig;color_selection}) and the spectral energy distribution (SED) fitting (Section \ref{sec;sed}), the template-fitting photometries are used (see \citealt{2013ApJS..206...10G} for more details).
Because the effective wavelengths of the NB209/NB2315 filters are slightly different from that of the $K$-band filter,
we estimate the continuum level at the wavelengths of the NB209/NB2315 filters by taking a linear interpolation between $H$-band and $K$-band magnitudes as follows,

\begin{eqnarray}
HK(\lambda=2.09 \mu \mathrm{m}) &=&0.8K+0.2H-0.015, \\
HK(\lambda=2.31 \mu \mathrm{m}) &=&1.2K-0.2H+0.011.
\end{eqnarray}

\noindent
Figure \ref{fig;emitter_selection} shows the $HK-$NB209 and $HK-$NB2315 color-magnitude diagrams for all the objects
that are brighter than the 5$\sigma$ limiting magnitudes of NB209 and NB2315, respectively.
For faint objects ($<2\sigma$) in $H$-band or $K$-band, the $2\sigma$ limiting magnitude is used to calculate colors.
\cite{1995MNRAS.273..513B} defined a significance of a magnitude excess in a certain NB by a parameter $\Sigma$ taking account of the fact that the fainter the NB flux is, the larger the photometric error is, as follows;

\begin{equation}
m_\mathrm{BB}-m_\mathrm{NB}=-2.5\log\left[1-\frac{\Sigma\sqrt{f_{1\sigma\mathrm{BB}}^2+f_{1\sigma\mathrm{NB}}^2}}{f_{\mathrm{NB}}} \right],
\end{equation}

\noindent
where $m$ and $f$ indicate a magnitude and flux density, respectively.
$\Sigma$=3 and $HK-NB>0.4$, which corresponds to the equivalent width cut in the rest frame of EW$_\mathrm{rest}\sim40$ \AA, are adopted to select NB emitters. 
These criteria are similar to those used in our previous NB survey \citep{2012MNRAS.423.2617T}.
The significance of $\Sigma=3$ corresponds to the limiting line fluxes of $1.5~(2.6)\times 10^{-17}$ erg s$^{-1}$ cm$^{-2}$, that can be transformed to 4 (10) $M_\odot$yr$^{-1}$ of dust-uncorrected SFRs \citep{1998ARA&A..36..189K}.
Based on the first criteria, 99 (58) objects are identified as NB209 (NB2315) emitters, respectively.

\begin{table*}
\begin{center}
\caption{Spectroscopic redshift, stellar mass and metallicities. \label{tab;spectroscopy}}
\begin{tabular}{lcccccc}
\hline
ID & R.A. & Decl. & $z_\mathrm{H\alpha}$ & $M_*$  & [N~{\sc ii}]/$H\alpha$ & 12+log[O/H] \\
& (J2000) & (J2000) & & ($\times10^{10} M_\odot$)  & & \\
\hline
SXDF-NB209-2   & 02 17 17.42 & $-$05 13 48.4 & 2.183 &  16.3 &  0.63 $\pm$ 0.16 &  \\
SXDF-NB209-4   & 02 17 04.99 & $-$05 12 28.4 & 2.191 &  24.8 &  $<$0.27 & $<$8.58 \\
SXDF-NB209-6   & 02 17 10.51 & $-$05 09 16.2 & 2.180 &   3.2 &  0.35 $\pm$ 0.10 & 8.64 $\pm$ 0.07 \\
SXDF-NB209-7   & 02 17 14.71 & $-$05 11 42.0 & 2.192 &   3.6 &  0.27 $\pm$ 0.05 & 8.58 $\pm$ 0.05 \\
SXDF-NB209-9   & 02 17 07.90 & $-$05 14 15.4 & 2.186 &  10.7 &  0.71 $\pm$ 0.22 &  \\
SXDF-NB209-11 & 02 17 07.85 & $-$05 12 26.6 & 2.191 &   3.1 &  0.36 $\pm$ 0.08 & 8.65 $\pm$ 0.05 \\
SXDF-NB209-12 & 02 17 07.61 & $-$05 09 17.3 & 2.180 &  21.5 &  0.29 $\pm$ 0.08 & 8.59 $\pm$ 0.07 \\
SXDF-NB209-16 & 02 17 13.49 & $-$05 10 05.5 & 2.180 &  18.1 &  0.78 $\pm$ 0.26 &  \\
SXDF-NB209-17 & 02 17 11.06 & $-$05 12 49.0 & 2.182 &  14.5 &  0.55 $\pm$ 0.14 & 8.75 $\pm$ 0.06 \\
SXDF-NB209-34 & 02 17 08.09 & $-$05 14 16.1 & 2.184 &   0.3 &  $<$0.11 & $<$8.35 \\
SXDF-NB209-43 & 02 17 12.12 & $-$05 12 51.5 & 2.182 &   1.2 &  $<$0.18 & $<$8.48 \\
SXDF-NB209-60 & 02 17 08.66 & $-$05 09 24.1 & 2.180 &   0.4 &  $<$0.33 & $<$8.63 \\
\hline
\end{tabular}
\end{center}
\end{table*}

The NB209 (NB2315) filter picks up not only a \ha line at $z=2.2~(2.5)$ but some other lines at different redshifts are contaminating such as \pa at $z=0.1~(0.2)$, \oiii at $z=3.2~(3.6)$ and \oii line at $z=4.6~(5.2)$. 
Although a photometric redshift technique is expected to be effective in discriminating HAEs at $z=2.2$ and $2.5$ from a few other choices, it is less accurate for some faint galaxies where photometric redshift errors are large.
Therefore, instead of using photometric redshifts, we define the best color selections to cull out the HAEs based on the spectroscopic sample of galaxies in the MOIRCS Deep Survey (MODS; \citealt{2011PASJ...63S.379K}).
Figure \ref{fig;color_selection} shows the color-color diagrams for the galaxies with spectroscopic redshifts.
The color combination is chosen so that it can capture the Balmer/4000 \AA ~break of galaxies at each redshift.
The $BiK$ and $iJK$ diagrams are found to effectively isolate $z\sim0.1$ emitters and $z\sim3.2$ emitters, respectively, from the HAEs at $z=2.2$.
As for the HAEs at $z=2.5$, the $BiK$ and $iHK$ diagrams are used to reject $z\sim0.2$ emitters and $z\sim3.6$ emitters.
The degrees of completeness and contamination in the color selections are estimated, based on the spectroscopically confirmed MODS sample.
98\% (91\%) of the MODS sample at $z\sim2.2$ ($z\sim2.5$) satisfy our color criteria and 5\% (5\%) of the MODS sample at other redshift are also included as contaminants.
After these color selections, 63 HAEs at $z=2.2$ and 46 at $z=2.5$ are finally identified.
While most of NB2315 emitters are HAEs at $z=2.5$, the NB209 sample includes about 30 \oiii emitters at $z=3.2$.

\subsection{MOIRCS spectroscopic follow-up}
\label{sec;spectra}

\begin{figure*}
\begin{center}
\includegraphics[width=0.8\linewidth]{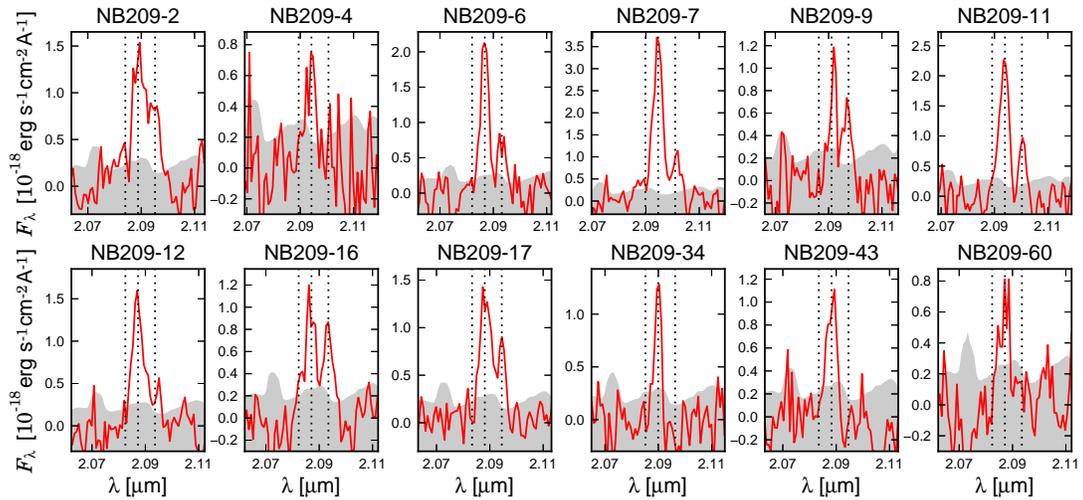} 
\caption{\ha line spectra of our twelve NB209 HAEs. Red lines show the observed spectra. Vertical dotted lines indicate the locations of \ha and [N~{\sc ii}]$\lambda_06548,\lambda_06583$ lines. Gray zones show the sky noise levels. Flux is calibrated by using a star in the 2-MASS catalog \citep{2006AJ....131.1163S},
although we do not rely on the flux calibration in this paper. \label{fig;spectra}}
\end{center}
\end{figure*}

In order to confirm that the HAEs, which will be selected in Section \ref{sec;target_selection}, are actually located at $z=2.2$, and to measure their spectral properties such as the line ratio of [N~{\sc ii}]/H$\alpha$,
we conducted a spectroscopic follow-up observation with MOIRCS on the multi-object spectroscopy (MOS) mode \citep{2006SPIE.6269E.148T} in 2012 October--November.
The HK500 low-resolution grizm, which covers the wavelength range of $\lambda=1.3--2.3~\mu$m, was used.
The slit width was 0\arcsec.8 and its length was 10-12\arcsec.
The total on-source exposure time was 220 minutes. 

Data reduction is conducted with the MOIRCS MOS Data Pipeline (MCSMDP; \citealt{2010ApJ...718..112Y}) in the following way. 
First of all, bad pixels and cosmic-rays are removed.
A sky frame is created in parallel to processing the object frames.
The object frames and the sky frame are then divided by the dome flat for flat-fielding. 
Distortion is corrected by using the database for the imaging mode of MCSRED. 
The wavelength calibration is performed by using the OH lines in the sky frame.
After a sky pattern is roughly removed by subtracting the adjacent dithered image (A--B subtraction) for each object frame, 
the residual background emission is subtracted by using the task ``background'' of IRAF. 
Finally, all frames are combined with weights in exposure times.
To calibrate the telluric absorption and the difference in sensitivity at each wavelength, the object frames are divided by the spectrum of a spectroscopic standard star (A1-type), and multiplied by the model spectrum \citep{2004astro.ph..5087C}.
For a slit spectroscopy, a slit loss must be properly corrected for in order to estimate a total flux.
However, the slit loss strongly depends on the size of a galaxy.
Therefore, in this work, we do not use the absolute flux derived from the
spectroscopy and discuss only the redshifts and the line ratios of the HAEs.

For the target selection of spectroscopy, we gave a high priority on objects with a bright \ha emission measured from the NB data.
With the MOIRCS, the NIR spectra of 13 HAEs at $z=2.2$ have been obtained.
We successfully detected \ha and [N~{\sc ii}] emission lines above $3\sigma$ from 12 and 8 objects, respectively.
Table \ref{tab;spectroscopy} lists their spectroscopic redshifts, and Figure \ref{fig;spectra} shows their spectra.
For only one object, no line is seen in its spectrum though the expected \ha flux density from our NB data is large enough to be detected.
The non-detection might be due to a large flux loss by the slit because it is a diffuse source.
In order to measure the spectroscopic redshifts and line ratios, 
the reduced spectra are fit by a three-component (\ha and [N~{\sc ii}]$\lambda6548,\lambda6583$) Gaussian
model, with the free parameters of redshift, line width, and flux density of each line.
The continuum is first subtracted off by fitting the spectra with a linear function
over the wavelength range of 2.08--2.1 $\mu$m.
The line ratio of [N~{\sc ii}]$\lambda6583$/[N~{\sc ii}]$\lambda6548$ is fixed to 3.0
\citep{2000MNRAS.312..813S}.
The sky noise is estimated from a Poisson error of the sky frame without a background
subtraction.

For SXDF-NB209-4, 34, 43 and 60, \nii lines are not detected.
SXDF-NB209-34 has been confirmed to be located at $z=2.2$ as its \oiii and \hb emission lines are detected.
Although we do not yet have conclusive evidence at this stage that the detected lines are \ha for SXDF-NB209-4, 43 and 60,
our spectroscopic follow-up observations do support that our criteria of identifying HAEs based on NB and BB data should work very well.
We hereafter regard all of the 12 sources with line detections as HAEs.

\section{Galaxy properties}

\subsection{SED fitting}
\label{sec;sed}

Multi BB photometries provide SEDs of individual galaxies, which give us valuable information of physical properties such as stellar mass, age, and amount of dust extinction.
We fit the photometric SEDs traced by 12 bands ($u, B, V, R, i', z', Y, J, H, K_s, 3.6~\mu \mathrm{m}, 4.5~\mu \mathrm{m}$) with the stellar population synthesis model of \cite{2003MNRAS.344.1000B}. We adopt the solar metallicity.
The SED fitting with a standard $\chi^2$ minimization procedure is applied by using $hyperz$ \citep{2000A&A...363..476B}.
The redshift is fixed to $z=2.19$ and $z=2.53$ for the NB209 and NB2315 HAEs, respectively.
We use the libraries with exponentially declining star formation histories: SFR$\propto e^{-t/\tau}$ with $\tau$=0.01, 0.02, 0.05, 010, 0.20, 0.50, 1.0, 2.0, and 5.0 Gyr.
Salpeter IMF is adopted for the estimation of stellar masses and SFRs.
The dust attenuation law for starburst galaxies \citep{2000ApJ...533..682C} is applied to the model spectra.
The best fitting template gives stellar mass, age, $A_V$ and $\tau$.
Although the estimated age and the amount of dust extinction by SED fitting are not very reliable due to a degeneracy, the stellar mass is a rather robust quantity if IMF is given.
Ideally, we should accurately measure the amount of dust extinction by the Balmer decrement method. 
However, \hb lines are generally too weak to be detected for high-$z$ galaxies, and we ought to rely on the values of dust extinction derived from the SED fitting.

\subsection{AGN contribution}
\label{sec;AGN}

Ideally, we can distinguish between star-forming galaxies and active galactic nuclei (AGNs) by measuring emission line ratios of [O~{\sc iii}]/\hb and [N~{\sc ii}]/\ha and plot them on the so-called BPT diagram \citep{1981PASP...93....5B}.
We detected both \hb and [O~{\sc iii}] emission lines for only one target (SXDF-NB209-34).
For this object, the line ratios of log ([O~{\sc iii}]/H$\beta$)=0.73 and log ([N~{\sc ii}]/H$\alpha$)$<-0.96$ indicate that this is a star-forming galaxy.
Unfortunately, neither \hb nor \oiii lines are detected for the other objects, and so we are obliged to make a diagnosis based solely on the line ratio of [N~{\sc ii}]/H$\alpha$.
[N~{\sc ii}]/H$\alpha$ is particularly sensitive to shock excitation or the presence of a hard ionizing radiation field around an AGN. 
\cite{2002ApJS..142...35K} put a theoretical upper limit on [N~{\sc ii}]/\ha ratio based on photo-ionization and stellar population synthesis models. For star-forming galaxies, the ratio of [N~{\sc ii}]/\ha should always be less than 0.6.
On the BPT diagram, Seyfert galaxies are often defined to have [O~{\sc iii}]/H$\beta>$ 3 and [N~{\sc ii}]/\ha $>$ 0.6, and low-ionization nuclear emission-line regions to have [O~{\sc iii}]/\hb $<$ 3 and [N~{\sc ii}]/\ha $>$ 0.6 \citep{2003MNRAS.346.1055K}.
Therefore, we regard the three HAEs (SXDF-NB209-2, 9 and 16) with [N~{\sc ii}]/H$\alpha>$ 0.6 as AGNs.

One HAE (SXDF-NB209-1) has been diagnosed as a narrow-line AGN on the basis of high-ionization ultraviolet emission lines such as C~{\sc iv} and [Ne~{\sc iii}] by the previous study \citep{2012MNRAS.421.3060S}.
The diagnostic with IRAC four-band photometries is also useful for identifying AGNs. 
The mid-infrared SEDs of AGNs are well fitted by a power law because the thermal emission by hot dust dominates the SEDs of host galaxies \citep{2007ApJ...660..167D}. 
\cite{2008ApJ...687..111D} find that a power-law selection of $\alpha<-0.5$ recovers the majority of high-quality AGN candidates. 
While the IRAC SEDs of most our HAEs are not fitted by a power-law, one HAE (SXDF-NB209-42) satisfied the criterion of $\alpha<$-0.5, suggesting that it hosts an AGN. 
In total, we have identified five AGN candidates.
In SXDF, deep X-ray data by $XMM-Newton$ are available \citep{2008ApJS..179..124U}, which enable us to identify AGNs.
However, none of our five AGN candidates are detected in X-ray, suggesting that they are obscured AGN by dust and gas.

\begin{figure}
\begin{center}
\includegraphics[scale=1.0]{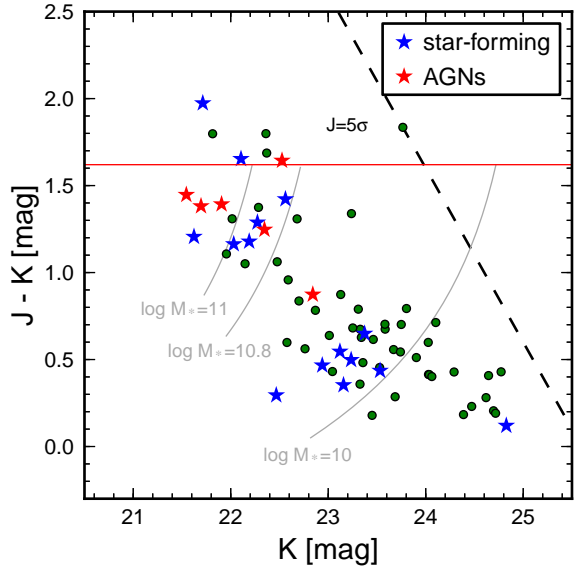}
\caption{The $J-K$ versus $K$ color-magnitude diagram for the HAEs at $z = 2.2$ in SXDF and GOODS-N field. Red stars show the identified AGNs and blue stars indicate the spectroscopically confirmed star-forming galaxies with [N~{\sc ii}]/\ha of $<$0.6. Green circles show the objects without spectra. The diagonal dashed line represents the detection limit of 5$\sigma$ in J-band. The red horizontal line indicates the expected location of a red sequence at $z=2.2$ if the formation redshift is $z_\mathrm{form}$ = 5, based on the \cite{1998A&A...334...99K} model. Gray lines denote the iso-stellar mass curves for $M_*= 10^{11}, 10^{10.8}$ and $10^{10} M_\odot$, which are estimated by the \cite{1998A&A...334...99K} model. We note that the actual stellar masses of individual galaxies are derived from the full SED fitting (Section \ref{sec;sed}).\label{fig;CM}}
\end{center}
\end{figure}

Figure \ref{fig;CM} shows the color--magnitude diagram for HAEs at $z=2.2$.
We also plot seven HAEs at $z=2.2$ in GOODS-N field, which are identified by our previous NB209 survey on MOIRCS and confirmed by the follow-up spectroscopy \citep{2011PASJ...63S.437T}.
One of them is an X-ray-detected AGN.
Galaxies are known to show a color bimodality; blue cloud galaxies and red sequence galaxies \citep{2003MNRAS.341...54K}. 
The HAEs are naturally supposed to be blue star-forming galaxies.
However, some of them are red in the rest-frame optical colors ($J - K >$ 1). 
Throughout this paper, "red" refers to a red color in the rest-frame optical.
The red HAEs tend to be more massive than normal blue emitters so that they are likely to be direct progenitors of massive galaxies seen today.
We find that at least 42\% (=5/12) of these red, massive HAEs with $M_*>10^{10.8} M_\odot$ are strongly influenced by AGNs while most of the blue, less massive ones are likely to be star-forming galaxies.
The dependence on stellar mass or color is consistent with the previous studies \citep[e.g.,][]{2009ApJ...699.1354Y,2010ApJ...720..368X}.

\begin{figure}
\begin{center}
\includegraphics[scale=1.0]{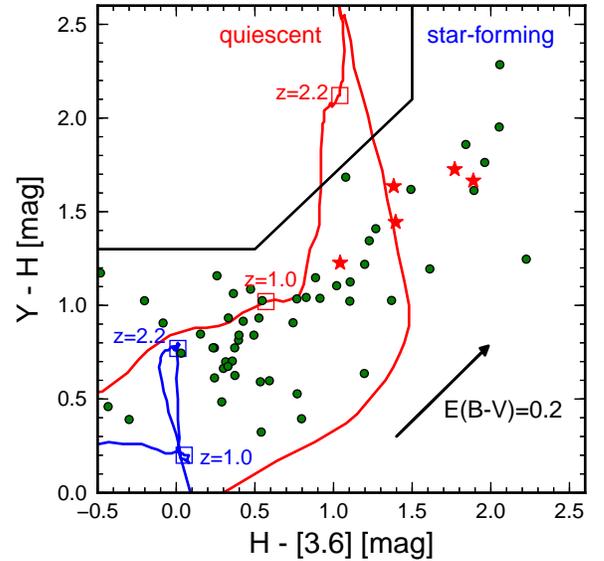}
\caption{The $Y-H$ versus $H-$[3.6] color-color diagram for the HAEs at $z=2.2$ in SXDF. Red stars denote the AGNs. Red and blue lines show a model track of a quiescent galaxy and a star-forming galaxy, respectively \citep{1998A&A...334...99K}. The arrow indicates the reddening vector corresponding to E(B-V)=0.2 \citep{2000ApJ...533..682C}.\label{fig;YHIRAC}}
\end{center}
\end{figure}

We investigate whether the galaxies hosting AGNs are quiescent or dusty star-forming based on the $Y-H$ versus $H-$[3.6] color-color diagram (Figure \ref{fig;YHIRAC}).
It is found that their colors are inconsistent with ones of the quiescent galaxy model and redden along with the dust reddening vector of star-forming galaxy model.
Therefore, they are likely to be mixed systems of dusty star-forming galaxies containing AGNs rather than quiescent galaxies.
Our results suggest that AGN activity is enhanced at the late stage of massive galaxy formation and possibly contributes to subsequent quenching of star formation.
An investigation of the fraction of AGNs for a larger, complete HAEs sample by additional follow-up spectroscopic observations is desperately needed..
In the following sections, we exclude these AGNs to concentrate on star-forming activities. 

\subsection{Star formation rates}
\label{sec;sfr}

It is standard for us to estimate SFRs of our HAEs with the following standard calibrations of \cite{1998ARA&A..36..189K};

\begin{eqnarray}
\mathrm{SFR}_{\mathrm{H}\alpha} [M_\odot \mathrm{yr}^{-1}] &=& 7.9\times 10^{-42}~L_\mathrm{H\alpha }  [\mathrm{erg\ s}^{-1}], \\
\mathrm{SFR}_{\mathrm{UV}} [M_\odot \mathrm{yr}^{-1}] &=&1.4\times 10^{-28}~L_\mathrm{UV}  [\mathrm{erg\ s}^{-1}\mathrm{Hz}^{-1}],
\end{eqnarray}

\noindent
where $\mathrm{SFR}_\mathrm{H\alpha}$ and $\mathrm{SFR}_\mathrm{UV}$ indicate dust-uncorrected SFRs based on \ha and rest-frame UV luminosities, respectively.
Since the composite UV spectrum is nearly flat over the wavelength range of 1500--2800 \AA, 
the UV luminosity is calculated by linearly-interpolating the $B$-, $V$- and $R_c$-band luminosities as,

\begin{eqnarray}
m_{1500}&=&0.7B+0.3V\ \ \ \mathrm{for\ HAEs\ at}\ z=2.2,\\
m_{1500}&=&1.2V-0.2R_c\ \ \mathrm{for\ HAEs\ at}\ z=2.5,
\end{eqnarray}

\begin{figure}
\begin{center}
\includegraphics[scale=1.0]{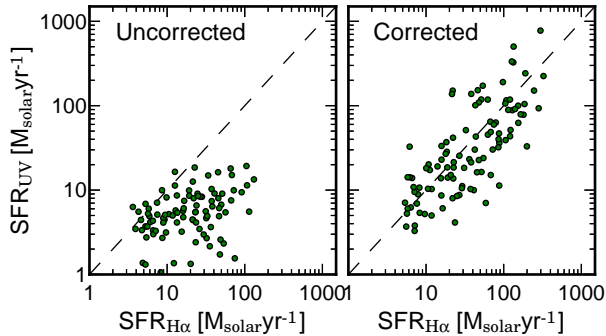}
\caption{$Left$: H$\alpha$-based SFRs (SFR$_\mathrm{H\alpha}$) versus UV-based SFRs (SFR$_\mathrm{UV}$) without dust-correction for the combined sample of HAEs at $z=2.2$ and $z=2.5$. $Right$: Same as the left panel, but with the dust extinction correction based on the SED fitting. \label{fig;Ha_UV}}
\end{center}
\end{figure}

\noindent
where $m_{1500}$ is the interpolated magnitude at the rest-frame wavelength of 1500 \AA.

The HAE surveys with NB filters allow us to measure the \ha fluxes from the flux densities in NB and BB, hence SFRs for all galaxies \citep{2010MNRAS.403.1611K}.
The NB flux density can be defined as $f_\mathrm{NB}=f_c+F_\mathrm{line}/\Delta _\mathrm{NB}$, where $f_c$ is the continuum flux density, $F_\mathrm{line}$ is the emission-line flux, and $\Delta$ denotes FWHMs of the filters.
The BB flux density is also defined as $f_\mathrm{BB}=f_c+F_\mathrm{line}/\Delta _\mathrm{BB}$.
Therefore, the line flux, continuum flux density, and equivalent width (EW) are calculated as,

\begin{eqnarray}
F_\mathrm{line}&=&\Delta _{\mathrm{NB}}\frac{f_{\mathrm{NB}}-f_{\mathrm{BB}}}{1-\Delta _{\mathrm{NB}}/\Delta _{\mathrm{BB}}},\\
f_\mathrm{c}&=&\frac{f_{\mathrm{NB}}-f_{\mathrm{BB}}(\Delta _{\mathrm{NB}}/\Delta _{\mathrm{BB}})}{1-\Delta _{\mathrm{NB}}/\Delta _{\mathrm{BB}}},~\mathrm{and}~\\
\mathrm{EW_{rest}}&=& \frac{F_\mathrm{line}}{f_\mathrm{c}}(1+z)^{-1}.
\end{eqnarray}

\noindent
Note that a \nii line also contributes to a NB209/NB2315 flux. 
\cite{2012MNRAS.420.1926S} estimated the [N~{\sc ii}] contamination for the full Sloan Digital Sky Survey (SDSS) sample as a function of total rest-frame EW and compute the polynomial approximation of the relationship between log([N~{\sc ii}]/H$\alpha$) (denoted as $f$) and log[EW$_\mathrm{rest}$([N~{\sc ii}]+H$\alpha$)] (denoted as $E$), which is presented as $f=-0.924+4.802E-8.892E^2+6.701E^3-2.27E^4+0.279E^5$. 
We use this polynomial function to correct for a \nii contamination to derive
a pure \ha flux. The average correction is [N~{\sc ii}]/H$\alpha$=0.16.

Figure \ref{fig;Ha_UV} shows the SFR$_{\mathrm{H}\alpha}$ versus SFR$_\mathrm{UV}$ diagram for the HAEs at $z=2.2$ and 2.5 with and without dust extinction correction (right and left panels, respectively). 
The dust uncorrected SFR$_\mathrm{UV}$ is not correlated to SFR$_\mathrm{H\alpha}$
(left panel). 
We use the amount of dust extinction derived from the SED fitting in Section \ref{sec;sed} and assume equal extinction between UV (stellar) and \ha (nebular) rather than $E(B-V$)$_\mathrm{stellar}$=0.44$E(B-V$)$_\mathrm{nebular}$ \citep{2000ApJ...533..682C}, because the SFR$_{\mathrm{H}\alpha}$ is significantly overestimated with respect to the SFR$_\mathrm{UV}$ in the latter situation \citep{2006ApJ...647..128E}.
Once we apply such corrections, the right panel of Figure \ref{fig;Ha_UV} shows a good correlation between the two measurements of SFRs from two independent indicators.
Hereafter, we use dust-corrected SFRs based on \ha luminosities, in order to define the main sequence of star-forming galaxies at $z>2$ presented in Section \ref{sec;main_sequence}.


\begin{figure}
\begin{center}
\includegraphics[scale=1.0]{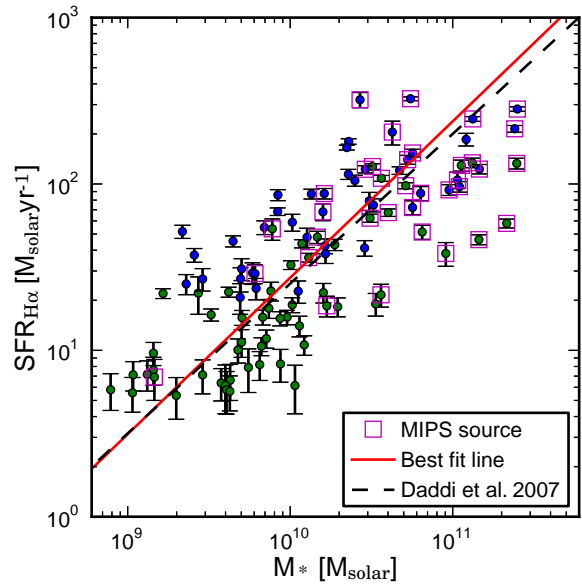}
\caption{Star formation rates of the combined sample of HAEs at $z=2.2$ (green) and $z=2.5$ (blue) plotted against stellar masses. We use the \ha-based SFRs with the dust extinction correction (Section \ref{sec;sfr}). The errors in SFRs are estimated from the photometric errors ($\sigma=1$). 
The red solid line indicates the best fit line to our data points (SFR=238$M_{11}^{0.94}$). 
The dashed line presents the main sequence of star-forming galaxies at $z\sim2$ defined by \cite{2007ApJ...670..156D}. Magenta squares indicate the MIPS 24 $\mu$m-detected sources. \label{fig;MS}}
\end{center}
\end{figure}

\subsection{Main sequence of star-forming galaxies}
\label{sec;main_sequence}

The $M_*-$SFR relation, named as the $main~sequence$ of star-forming
galaxies, is conveniently parameterized as

\begin{equation}
\mathrm{SFR}=\alpha \times M_*^\beta.
\label{eq;ms}
\end{equation}

\noindent
Based on the fact that the majority of star-forming galaxies are located on this sequence,
it must have essential information on how star formation in galaxies are generally
regulated, and it must be playing a vital role in galaxy evolution.
In spite of its vital importance, the values of $\alpha$ and $\beta$ are still quite
uncertain especially at high redshifts. This is because the results are highly dependent
both on the sample selection and on the SFR indicator that is used.

Due to our relatively un-biased sample of star-forming galaxies and the robust measure of  SFRs based on H$\alpha$, we are able to investigate such a relationship for star-forming galaxies at $z=2.2$ and $2.5$.
Figure \ref{fig;MS} shows the SFR--$M_*$ diagram for the HAEs.
The star forming activities of HAEs at $z=2.5$ seem to be systematically higher than those at $z=2.2$ at the fixed stellar mass.
It is not clear at this moment whether such difference is due to an intrinsic evolution of galaxies with time, or an environmental dependence (cosmic variance), or simply due to small number statistics. 
We will investigate this potentially interesting result in greater detail in a future paper.
In this paper, we do not consider this possible difference further, and we fit a relation for the combined sample of $z=2.2$ and $2.5$ with a stellar mass cut of $M_*>10^{9.5}~M_\odot$ since our HAE sample becomes significantly incomplete below it.
Also, we do not use massive galaxies with $M_*>10^{10.8}~M_\odot$, in which the AGN fraction could increase (Section \ref{sec;AGN}), for the fit.
The best fit line is SFR=238$M_{11}^{0.94}$, where $M_{11}$ shows the stellar mass in the unit of $10^{11}~M_\odot$ ($M_\mathrm{11}$=$M_*/10^{11}~M_\odot$), and its scatter is 0.3 dex.
\cite{2007ApJ...670..156D} gives SFR=200$M_{11}^{0.9}$ as the main sequence of color-selected star-forming galaxies at $z\sim2$.
The best fitted main sequence is well suited to that defined by \cite{2007ApJ...670..156D} although they estimated SFRs by using UV and IR luminosities rather than \ha emission lines, and the sample selection is also different from ours.
Hereafter, we use our fitting result for the main sequence.

\subsection{Mass--metallicity relation}
\label{sec;metal}

The [N~{\sc ii}]/\ha line ratios can also be used as a crude estimator of gaseous metallicity
in H{\sc ii} regions of star-forming galaxies.
A metallicity is one of the most important properties for understanding galaxy evolution
because it imprints the past star-formation histories.
If a \ha emission line is free from AGN contribution, we are able to measure
its oxygen abundance with the line ratio of
N2=log([N~{\sc ii}]$\lambda$6583/H$\alpha$); 12+log(O/H)=8.9+0.57$\times$N2 \citep{2004MNRAS.348L..59P}. 

\begin{table*}
\begin{center}
\caption{Properties of HAEs in GOODS-N field\label{tab;spectroscopy_GOODS}}
\begin{tabular}{lcccccc}
\hline
ID & R.A. & Decl. & $z_\mathrm{H\alpha}$ & $M_*$  & [N~{\sc ii}]/$H\alpha$ & 12+log[O/H] \\
& (J2000) & (J2000) & & ($\times10^{10} M_\odot$)  & & \\
\hline
MODS-487     & 12 35 58.32 & 62 11 55.7 & 2.176 &   1.7 &  $<$0.33 & $<$8.63 \\
MODS-7651   & 12 37 01.97 & 62 15 50.0 & 2.188 &   0.9 &  $<$0.07 & $<$8.24 \\
MODS-7773   & 12 36 56.74 & 62 16 43.7 & 2.190 &   5.9 &  $<$0.26 & $<$8.57 \\
MODS-7889   & 12 36 52.90 & 62 17 26.5 & 2.187 &   0.5 &  $<$0.09 & $<$8.30 \\
MODS-9441   & 12 37 18.22 & 62 16 53.4 & 2.196 &   1.3 &  $<$0.07 & $<$8.24 \\
MODS-10522 & 12 37 25.54 & 62 19 09.8 & 2.192 &   0.7 &  $<$0.13 & $<$8.39 \\
\hline
\end{tabular}
\end{center}
\end{table*}

In the local universe, it is well known that the metallicity is correlated to the stellar mass \citep{2004ApJ...613..898T}. Such mass--metallicity relations were recently investigated even at $z>1$ by many authors.
For example, \cite{2006ApJ...644..813E} obtain the mass--metallicity relation with the stacked spectra of 87 rest-frame UV-selected star-forming galaxies at $z\sim2.2$ and found that their metallicities are systematically lower than those of local galaxies with the same stellar mass.
On the other hand, \cite{2009ApJ...691..140H} and \cite{2010ApJ...715..385O} report that optical-NIR color selected star-forming galaxies at $z\sim2$ show much higher metallicities 
which are almost comparable to those of local galaxies.
Since this difference is likely to be attributed to the difference in the sample selection,
it is important to investigate the mass--metallicity relation further with a larger, relatively
unbiased sample of star-forming galaxies such as HAEs.

\begin{figure}
\begin{center}
\includegraphics[scale=1.0]{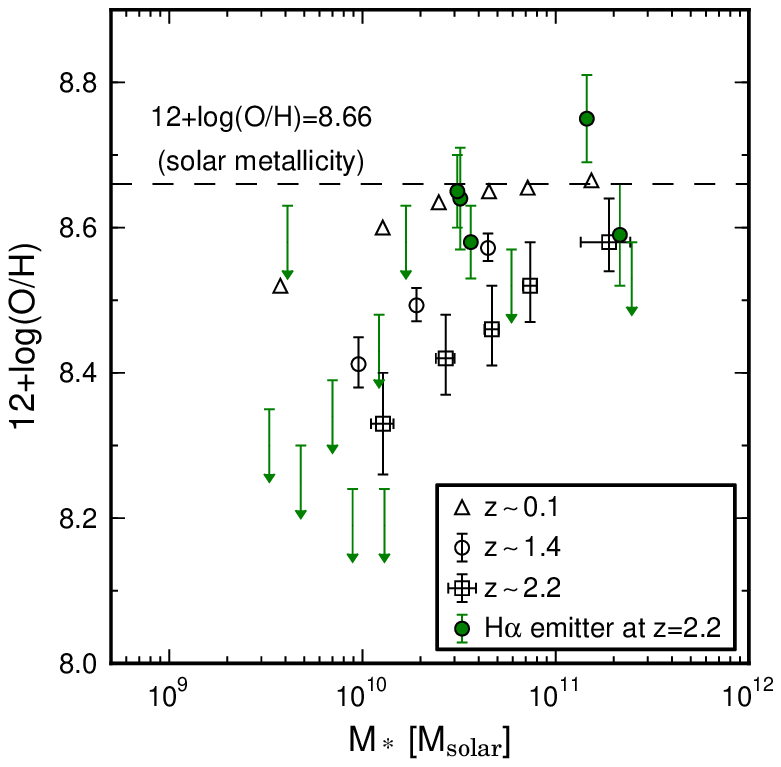}
\caption{Stellar mass versus metallicity relation for the HAEs at $z=2.2$ (green circles) in SXDF and in GOODS-N (\citealt{2011PASJ...63S.437T}). 
The upper limit points correspond to the 1$\sigma$ limiting flux of \nii line. Open squares and open triangles show UV-selected galaxies at $z\sim2.2$ \citep{2006ApJ...644..813E} and local galaxies at $z\sim0.1$ \citep{2004ApJ...613..898T}, respectively. The horizontal dashed line indicates the solar abundance of 12+log(O/H)=8.66 \citep{2004A&A...417..751A}. \label{fig;MZ}}
\end{center}
\end{figure}

Figure \ref{fig;MZ} presents the mass--metallicity relation for the spectroscopically confirmed HAEs at $z=2.2$ (Section \ref{sec;spectra}).
Note that our targets of spectroscopic observations are biased to objects with a large line flux.
For the objects whose \nii lines are not detected, we adopt 1$\sigma$ flux as the upper limit.
We also plot the sample of HAEs at $z=2.2$ in GOODS-N field, which are used in Section \ref{sec;AGN} \citep{2011PASJ...63S.437T}.
Table \ref{tab;spectroscopy_GOODS} lists the properties for six star-forming galaxies, excluding one X-ray detected AGN, in GOODS-N field.
Our data points for the HAEs are broadly consistent with the median relationship presented by \cite{2006ApJ...644..813E} for $z\sim2.2$, although including some upper limit points.
While some massive HAEs with \nii line detection are as metal rich as local galaxies at $z\sim0.1$ \citep{2004ApJ...613..898T},
the metallicities of less massive HAEs are systematically lower, and their dispersion is large at a given stellar mass from one sample to another.
Because our \ha selection by NB imaging provides us with a relatively unbiased sample of star-forming galaxies irrespective of their stellar populations, 
the variation we see here may represent the intrinsic scatter of the relationship.

\section{Discussion}

\subsection{Two modes of dusty star-forming galaxies}
\label{sec;dustiness}

In Figure \ref{fig;MS}, we find that the HAEs at the massive end of the main sequence tend to be detected at MIPS 24 $\mu$m, indicating that they are dusty star-forming galaxies.
However, for the galaxies below the detection limit of the MIPS data, we can not evaluate how dusty they are from the MIPS data alone.
To investigate a possible difference in dusty nature between the normal mode of star formation on the main sequence and the burst mode of star formation above the main sequence,
we define the ``dustiness'' index with the following equation,

\begin{eqnarray}
\mathrm{dustiness}&\equiv&\frac{\mathrm{SFR}_\mathrm{H\alpha,dust-uncorrected}}{\mathrm{SFR}_\mathrm{UV,dust-uncorrected}}.
\end{eqnarray}

\begin{figure}
\begin{center}
\includegraphics[scale=1.0]{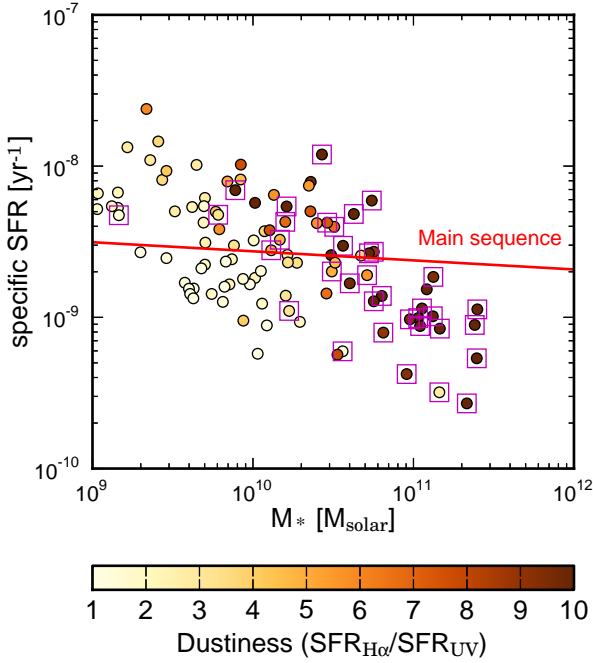}
\caption{Specific star formation rates versus stellar masses for the HAEs. Color scale indicates the dustiness (=SFR$_\mathrm{H\alpha}$/SFR$_\mathrm{UV}$). The red line shows the main sequence of HAEs at $z>2$. Magenta squares indicate the MIPS 24 $\mu$m-detected sources.\label{fig;sSFR}}
\end{center}
\end{figure}

\noindent
For dusty star-forming galaxies, the dustiness index should be high because the \ha emission line is much less attenuated by dust extinction than the UV light.
Most of the MIPS-detected HAEs actually show high values of dustiness ($>5$).
We could in principle estimate the amount of dust extinction ($A_V$) directly from the dustiness index by assuming the extinction curve such as \cite{2000ApJ...533..682C}: $A_V=[2.5\log (\mathrm{SFR}_\mathrm{H\alpha}/\mathrm{SFR}_\mathrm{UV})+0.4]/1.07$.
However, we choose to use the dustiness index because of the uncertainties in the extinction law.

\begin{figure}
\begin{center}
\includegraphics[scale=1.0]{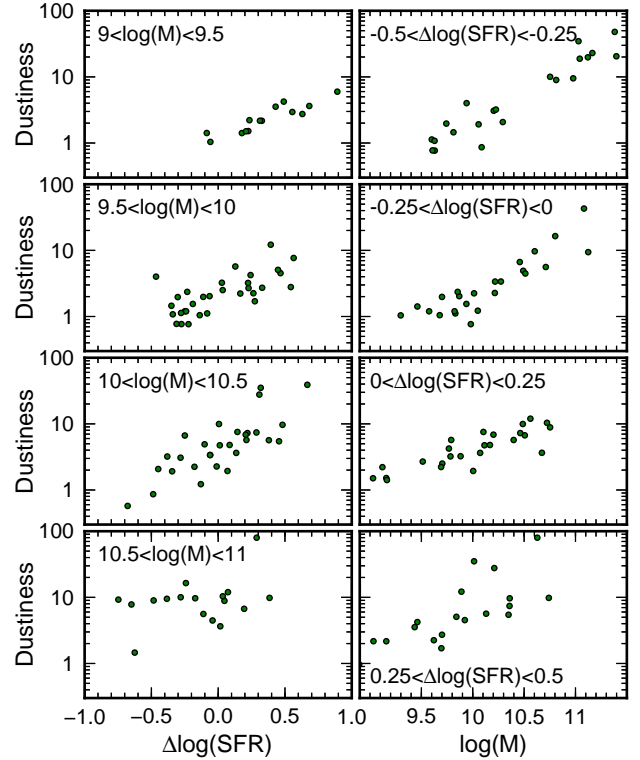}
\caption{Dustiness index (=SFR$_{\mathrm{H}\alpha}$/SFR$_{\mathrm{UV}}$) is plotted against the offset from the main sequence at fixed stellar mass (left row), and that plotted against the stellar mass at fixed $\Delta$log(SFR) (right row).\label{fig;M_sSFR_dustiness}}
\end{center}
\end{figure}

Figure \ref{fig;sSFR} plots the sSFRs as a function of stellar mass for our HAEs.
It seems that the dustiness depends on two parameters: offset from the main sequence in SFR, $\Delta$log(SFR)=log(SFR)-log(SFR$_\mathrm{MS}$), and stellar mass.
The HAEs with enhanced SFRs relative to the main sequence galaxies tend to have dusty formation.
On the other hand, the dustiness is also clearly high at the massive end despite declined star formation activities (low sSFRs).
These two interesting trends have been pointed out also by \cite{2011ApJ...742...96W}, who find that SFR$_\mathrm{IR}$/SFR$_\mathrm{UV}$ is correlated with both star formation activities and stellar mass.
We investigate the dependence of dustiness on two parameters separately by fixing one parameter and allowing the other parameter to change, and vice versa (Figure \ref{fig;M_sSFR_dustiness}).
For galaxies with $M_*<10^{10.5}~M_\odot$, the dustiness is closely correlated with $\Delta$log(SFR).
This result may give us a hint on why the star formation activity is boosted in the galaxies with large $\Delta$log(SFR).
In the case of clump migration \citep{2010MNRAS.404.2151C,2012MNRAS.422.1902I} and gas rich mergers, star forming regions are likely to be concentrated toward galaxy centers due to an inflow of gas that has lost its angular momentum.
Such compact dusty star forming regions would lead to a large dust attenuation.
We may therefore call such origin of dusty galaxies ``starburst mode''.

On the other hand, the dustiness seems to be constantly high regardless of the level of star formation activities for the massive galaxies with $M_*>10^{10.5}~M_\odot$.
This suggests that stellar mass also contributes to the dust obscuration.
In fact, at fixed $\Delta$log(SFR), the dustiness seems to increase along the main sequence from low mass to high mass ends.
Such dependence indicates an additional process to increase dust extinction as stellar mass is increased.  Star forming galaxies continue to grow in stellar mass and move up along the main sequence as more and more new stars are formed. During the growth, chemical evolution must progress in the galaxies and become progressively more metal rich. Since the dust production is strongly correlated with metallicity, 
galaxies would become dustier as a result.
Therefore, \ha emission lines and UV photons radiated from the ionized gas and massive stars, are strongly attenuated by dust, and they would appear as massive, dusty galaxies.
In fact, in the local universe, it is found that the amount of dust extinction roughly correlates
with stellar mass \citep{2010MNRAS.409..421G,2010MNRAS.405.2594G,2012MNRAS.420.1926S}.
We may call such origin of dusty galaxies as ``metal-enrichment mode''.

In summary, we propose that the two modes/origins of dusty star-forming
galaxies at $z\sim2$, are the starburst mode and the metal-enrichment mode.

\subsection{Origin of the metallicity variation at fixed stellar mass}
\label{sec;gas}

We find that the metallicities of HAEs at $z>2$ are significantly lower compared to local galaxies at the fixed stellar mass (Section \ref{sec;metal}).
It is not clear, however, why galaxies with the same stellar mass show such variations in metallicity despite of the fact that they have produced the same amount of stars in the past.
Recently, a fundamental relationship among three parameters, namely, mass, SFR, and metallicity, was presented, which can account for the source of scatter and the cosmic evolution of the mass--metallicity relation \citep[e.g.,][]{2010A&A...521L..53L,2012ApJ...761..126N}.
\cite{2010MNRAS.408.2115M} find that the metallicity is also correlated with SFR at fixed mass,
and the scatter is significantly reduced by introducing SFR as a second parameter.
Similarly, \cite{2012PASJ...64...60Y} demonstrate that galaxies with larger SFR have lower metallicity, with 71 star-forming galaxies at $z\sim1.4$.

\begin{figure}
\begin{center}
\includegraphics[scale=1.0]{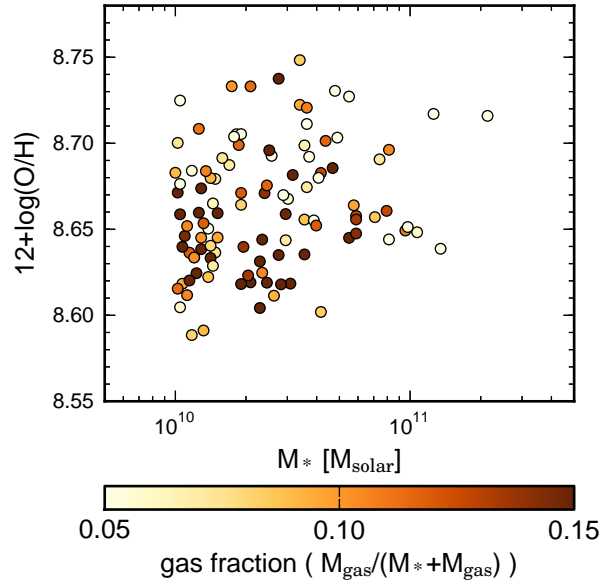}
\caption{Mass--metallicity relation for the star-forming galaxies at $0.025<z<0.05$ \citep{2011MNRAS.415...32S}. Color scale indicates the gas mass fraction of galaxies. \label{fig;MZ_GASS}}
\end{center}
\end{figure}

In order to investigate the effect of remaining molecular gas in the mass--metallicity relation,
we present the mass--metallicity relation for the SDSS sample at $0.025<z<0.05$ whose CO emission lines are detected in Figure \ref{fig;MZ_GASS}.
Here we use the spectroscopic SDSS data (DR7; \citealt{2009ApJS..182..543A}) and the catalog of CO Legacy Data base for the GASS (GALEX Arecibo Sloan Digital Sky Survey) survey \citep[COLD GASS;][]{2011MNRAS.415...32S}.
Molecular gas masses are derived from CO luminosities, with the conversion factor of 4.35 $M_\odot$/(K km s$^{-1}$pc$^{-2}$) for normal star-forming galaxies and 1.0 $M_\odot$/(K km s$^{-1}$pc$^{-2}$) in systems with high infrared luminosities ($L_\mathrm{IR}>10^{11}L_\odot$) and warm dust temperatures ($S_{60\mu\mathrm{m}}/S_{100\mu\mathrm{m}}>0.5$) \citep{2012ApJ...758...73S}.
The metallicities of galaxies at a given stellar mass show some scatter, but there is a clear trend that the galaxies with larger gas mass fraction, $M_\mathrm{gas}/(M_*+M_\mathrm{gas})$, tend to have lower metallicities, indicating that the gaseous metallicity is well correlated to gas mass and its fraction at fixed stellar mass in the local universe.
\cite{2013MNRAS.433.1425B} also find a dependence of metallicity on HI gas mass, supporting our result.
Galaxies with a large gas reservoir have a potential to evolve into more massive and metal rich galaxies by subsequent star formation, which would then be eventually located on the more massive side of the local mass--metallicity relation.
While gas fractions of local star-forming galaxies are distributed in a relatively narrow range (5\%--15\%), high$--z$ galaxies are more gas rich \citep{2010Natur.463..781T,2010ApJ...713..686D}. 
Therefore, the dependence of metallicity on the gas mass fraction would become more significant and visible.
In order to reveal the origin of the metallicity variation, we need to measure the amount of remaining cold gas and its fraction for a statistical sample of star-forming galaxies at $z>2$ with, eg.,\ ALMA.
We can then confirm the hypothesis that the metallicity variation at a given stellar mass is due to the scatter in the gas mass fraction.

\section{Summary}

We have conducted NB imaging and spectroscopic surveys of HAEs at $z=2.2$ and $z=2.5$ in SXDF, using MOIRCS on Subaru Telescope. Our survey has identified 109 HAEs at $z>2$ in total over a total of 180 arcmin$^2$ area.
We have confirmed probable \ha emission lines for 12 out of 13 targets by the spectroscopic follow-up observations. 
Therefore, our technique of searching for HAEs based on the excess fluxes in NB and multi-color selection, is proven to be a robust and efficient method. 
Based on this unique, clean sample of star-forming galaxies at $z\sim2$,
we have investigated their global properties in this paper.

\begin{enumerate}

\item The diagnostics based on the [N~{\sc ii}]/H$\alpha$ line ratio, high-ionization ultraviolet emission lines, or X-ray detections, shows that about 42\% of the red, massive HAEs with $M_*>10^{10.8}~M_\odot$ contain AGNs, implying that the AGN feedback may be contributing to quenching star formation in these massive systems.
To further investigate the effects of AGNs on the evolution of star-forming galaxies, we need to make a systematic, extensive spectroscopic observations over a large sample that spans a wide range in stellar mass.

\item The HAEs at $z=2.2$ and $z=2.5$ exhibit a well defined main sequence represented by SFR=238$M_{11}^{0.94}$. 
We find that the dustiness index (SFR$_{\mathrm{H}\alpha}$/SFR$_{\mathrm{UV}}$) of star-forming galaxies is dependent on two parameters: offset from the main sequence, $\Delta$log(SFR), and the stellar mass.
Galaxies with high SFRs with respect to the main sequence tend to have high dustiness.
On the other hand, massive star-forming galaxies also tend to be dusty,
probably because they are more metal rich and contain larger amounts of dust.
Although it is widely recognized that dusty galaxies such as SMGs are merger-driven starburst
galaxies \citep{2010ApJ...724..233E},
our result suggests that some dusty HAEs are not necessarily indicative of such populations and they can be more like metal rich normal star-forming galaxies.

\item The metallicities of HAEs are roughly consistent with the typical values of the UV-selected galaxies at $z\sim2.2$ \citep{2006ApJ...644..813E}.
Massive HAEs at $z>2$ have already evolved into a metal rich system to the same level as the local star-forming galaxies, while
the metallicities of less massive galaxies are systematically lower, and their dispersion is large.
The SDSS sample at $z\sim0.1$ with CO detections suggests that the scatter in the mass--metallicity relation originated from the different amounts of molecular gas fraction.
Gas rich galaxies tend to have lower metallicities at fixed stellar mass compared to gas poor ones.
Therefore, at $z>2$ when the gas mass fraction of galaxies is larger ($\sim40$\%) and the scatter
would also be large, the metallicities at a given stellar mass would naturally show a large scatter as observed.
The measurements of cold gas with ALMA and JVLA would reveal the dependence of metallicities on gas mass fraction at $z>2$ directly for the first time, and this may lead us to establish a new fundamental metallicity relation among three parameters, namely, stellar mass, metallicity, and gas mass fraction.

\end{enumerate}

This paper is based on data collected at Subaru Telescope, which is operated by the National Astronomical Observatory of Japan. 
We thank the Subaru telescope staff for their help in the observation. 
This work has made use of the Rainbow Cosmological Surveys Database, which is operated by the Universidad Complutense de Madrid (UCM), partnered with the University of California Observatories at Santa Cruz (UCO/Lick,UCSC).
Funding for the SDSS and SDSS-II has been provided by the Alfred P. Sloan Foundation, the Participating Institutions, the National Science Foundation, the U.S. Department of Energy, the National Aeronautics and Space Administration, the Japanese Monbukagakusho, the Max Planck Society, and the Higher Education Funding Council for England. The SDSS Web Site is http://www.sdss.org/.

We thank the anonymous referee who gave us many useful comments, which improved the paper.
K. T. thank Dr. Kazuhiro Shimasaku, Professor Masashi Chiba, Dr. Nobunari Kashikawa, Dr. Kentaro Motohara, Dr. Masami Ouchi, and Professor. Masanori Iye for useful discussions and comments.
K.T. and Y.K. acknowledge the support from the Japan Society for the Promotion of Science (JSPS) through JSPS research fellowships for young scientists.  
T.K. acknowledges the financial support in part by a Grant-in-Aid for the Scientific Research (Nos.\, 18684004, 21340045, and 24244015) by the Japanese Ministry of Education, Culture, Sports, Science and Technology.

\bibliographystyle{apj}
\bibliography{tadaki_2013}

\end{document}